# On the Double Natured Solutions of the Two-Temperature External Soft Photon Comptonized Accretion Disks

**Cesar Meirelles Filho***

Instituto de Astronomia, Geofisica e Ciencias Atmosfericas, Universidade de Sao Paulo, Sao Paulo, Brazil

Abstract: We have analyzed pair production in the innermost region of a two-temperature external soft photon comptonized accretion disk. We have shown that, if the viscosity parameter is greater than a critical value $\alpha_c$ , the solution to the disk equation is double valued: one, advection dominated and the other, radiation dominated. When $\alpha \leq \alpha_c$ , the accretion rate has to satisfy $\dot{m}_1 \leq \dot{m} \leq \dot{m}_2$ in order to have two steady state solutions. It is shown that these critical parameters $\dot{m}_1$, $\dot{m}_2$ are functions of r, $\alpha$, and $\theta_e$ , and $\alpha_c$ is function of r and $\theta_e$. Depending on the combination of the parameters, the advection dominated solution may not be physically consistent.

## INTRODUCTION

The idea that pair production could occur in astrophysical plasmas is not an old one. It dates from a pioneering paper by Bisnovatyi-Kogan, Zel' dovich and Sunyaev 1971[2]. In their study, these authors have considered the production of pairs through particle-particle interactions and have discovered that, under these conditions, there is a maximum temperature above which productions overwhelms annihilation, equilibrium is no longer possible and proton density approaches infinity. Soon, after BKZS's[2] conclusions on pairs on relativistic plasmas, a lot of works followed (Stoeger 1977[15]; Pozdnyakov, Sobol and Sunyaev 1977[13]: Liang 1979[6]) which failed to confirm the existence of this maximum temperature. However, though considering different processes for pair production, these works have focused on a rather limited range of the parameter space. A full understanding of BKZS's[2] results was only possible after the work of Lightman 1982[8], who showed the limitations of BKZS's[2] results. Accordingly, these results only hold for very small photon densities, or, equivalently, very small scattering depths. Much before attaining the maximum temperature, photon-photon processes will dominate pair production. In that regime, the relation between pair production and annihilation is quite different, not allowing for a critical temperature as is the case of particle-particle interaction. As a matter of fact, Lightman [8]has shown the existence of two branches of solution for a given ($\tau_p$, $T_e$): one, the low $n_+$, with particle-particle dominating pair production, and the other, the high $n_+$, with photon-photon interactions dominance in pair production. In the upper branch, the high $n_+$, the specific heat is negative, and as the heating rate and the luminosity increases, the temperature decreases; in the lower branch, the low $n_+$, as the heating rate increases, the temperature increases and eventually reaches a $T^c$, where both branches merge. For a further increase in the heating rate, the behavior follows that of the high $n_+$ branch, with temperature decreasing till the plasma becomes effectively thick. These findings of Lightman[8] have been confirmed independently by Svensson 1982[17], 1984[18].

When these results are applied to flows in accretion disks, the role of the temperature on the existence of equilibrium production-annihilation is now played by the accretion rate. If flows in accretion disks have conditions such that the dynamical time is shorter than the characteristic time for the ions to transfer energy to the electrons, radiative cooling will be inefficient and most of the energy will be advected with the ions. A two-temperature disk will follow, with both $T_i$ and $T_e$ close to their virial values, with $T_i >> T_e \sim m_e c^2$ , close to the inner radius. One, then, should expect a significant pair production in these flows. Kusunose and Takahara 1988[9] have shown that, for accretion rates below a critical one, there exist two branches of solution under equilibrium production-annihilation: one of them, the high pair, has z>1, and the other, the low pair solution, has z<1. For accretion rates greater than the critical one, no equilibrium solution is possible. An interesting result from these works is that the existence of a critical accretion rate is independent of pairs being created by particle-particle, particle-photon, or photon-photon processes, as long as the photons are internally produced. When photons are externally produced, the rate of pair creation is no longer non linear in the particle density.

As far as important issues like the number of solutions accessible to the disk for given physical conditions, the role of the accretion rate and viscosity in determining steady state solutions for accretion disks with pairs produced by internal photons are concerned, reasonable agreement is achieved by several authors(citation). However, if one is concerned with apir production by external photons, the situation is less clear, and the results are even discrepant. Tritz and Tsuruta 1989[19] and White and Lightman 1989[21] obtain rather similar results for disks with pairs produced by external soft photons. According to them, pairs produced in that way do not substantially modify the disk, the density of pairs being negligibly small. Since they failed to find a critical accretion rate, pair production-annihilation equilibrium is always possible. Their solution are not self-consistent in the sense that the ion temperature exceeds its virial value and the disk scale height is larger than the radial distance. Kusunose and Takahara 1990[10], on the other hand, relaxing the imposition of setting the y Kompaneetz parameter equal to one, have studied disks with pair production by external soft photons, with different soft photons input. They found a crtical accretion rate and pair densities smaller than the proton density. The critical accretion rate is proportional to the viscosity parameter and depends on the soft photon energy.

They also have found a unique solution to the disk equations. However, their result is only numerical and they have exploited only a small range for the soft photon energy, and they have not included advection in their energy equation. The effect of advection was later included by Kusunose and Mineshige 1996[11]. They, however, considered only internal soft photons produced by bremsstrahlung. Misra nad Melia 1995[12] have also considered pair production by soft external soft photons and internal bremsstrahlung photons. They have included a term of energy cooling due to photon escape, but they have not included advection.

In recent years, a large amount of observational data on compact X-ray and γ-ray sources are highly indicative of electron temperature close to the electron rest mass and ion temperature close to its virial value (citation) and, in the Inter Stellar Medium, close to some sources, there evidences of pair annihilation features (citation). This, besides being suggestive of the presence of pairs, implies not the importance of advective cooling, due to the increase of the scale height as compared to the radial distance, but as well that the inner region, where the disk has thickened, will be subject to a large amount of soft photons, irradiated from the outer disk.

If the picture we have about these sources, surrounded by accretion disks, is correct, the irradiation of the inner region is unavoidable. Therefore, it will be very useful to revisit disks with pairs produced by external soft photons. We, then, propose, in this paper, to undertake this task. Our special emphasis will be:

- The search for multiple solutions for soft photon comptonized two temperature accretion disks, with photons supplied by an external source;
- The search for critical values for both the accretion rate and the viscosity parameter;
- The search for conditions under which the solutions are physically consistent;
- The search for maximum electron temperature at each radius.

The outline of this paper is as follows. In§ I we present our model of accretion disk, together with some assumptions. In §II we explain the units and obtain the disk equations. In §III we obtain the solution to the disk equations. In § IV we obtain the critical parameters. In § V the conclusions are presented.

## I.ABOUT THE MODEL

The model of accretion disk we shall be considering is a soft photon Comptonized two-temperature accretion disk, with pair production by photon-photon interaction. The photons are externally produced and, after impinging the disk, they are upscattered in energy. The region we are most interested is the innermost one, where the disk has thickened. Most of the standard assumptions will be kept: keplerian velocity, stress tensor ~αP, α being the viscosity parameter and P the pressure, hydrostatic equilibrium in z-direction, particles obey a Maxwell-Boltzmann distribution. The energy transport will be radiative in z-direction, perpendicular to the plane of the disk, and advective in the radial direction. The advection of energy will be treated locally. The coupling between protons and electrons will only through Coulomb collisional energy exchange.

For the radiation field, we shall follow the treatment of Svensson 1984[18], and Zdziarski 1985[20], and assume it may be represented by an inverse power law with an exponential cutoff at $KT_e$ plus a Wien bump.

For the lepton energy equation we shall follow the Comptonization treatment as given by Zdziarski 1985[20], which is equivalent to to obtain a relation between the spectral index γ, the electron temperature and the probability of a power law photon into the region $x(=h\nu/ KT_e) \geq \theta_e$ $(KT_e /m_e c^2)$, where the Wien spectrum is formed.

As usual, it is assumed that heating of the disk by viscous processes locally balances cooling.

## II.DISK EQUATIONS

Throughout this article we shall adopt the following set of units: radial distance and the disk scale height are expressed in units of $R_S=2GM/c^2$, where $R_S$ is the Schwarzschild radius; electron temperature $\theta_e$ , and proton temperature $\theta_p$ in units of $m_e c^2$; M is the mass of the central object in units of 10 solar masses; $\dot{m}$ is the accretion rate in units of $L_E/c^2$ , where $L_E$ is the Eddington luminosity; z, the pair density will be expressed,in units of the electron number density $N_e$. In the following we shall adopt the same procedure of Björnsson and Svensson 1992[3], and Björnsson et al. 1996[4], and employ, as much as possible, as variables, the pressure P in units of rest mass energy density, the compactness parameter ℓ , or radiative flux in units of $4\pi m_e c^3/(3\sigma_T H)$, and the Thomson scattering depth, $\tau_p= \sigma_T N_e H$, where $\sigma_T$ is the Thomson cross section for electron scattering, and H is the disk scale height.

Using these variables, the disk structure equations, hydrostatic equilibrium, conservation of angular momentum and energy balance are written, respectively, as

$$P = \frac{1}{2}\frac{H^2}{r^3}\ , \tag{1}$$

$$P = \frac{1}{2^{1.5}\alpha\tau_p}\square r^{-1.5}S(r), \tag{2}$$

$$\ell = \frac{\pi}{2}\frac{m_p}{m_e}Ar^{-2}(\frac{H}{r})\dot{\mathrm{m}}\,S(r), \tag{3}$$

Now, defining η in the same way as Björnsson and Svensson 1996, i.e.,

$$\eta = \square r^{-1.5}S(r), \tag{4}$$

And substituting the scale height by the pressure, we obtain

$$P\tau_p = \frac{1}{2^{1.5}}\frac{\eta}{\alpha}, \tag{5}$$

And

$$\square = \frac{\pi}{\sqrt{2}}\frac{m}{m}A\eta P^{0.5},$$

(6)

Where

$$A = 1 - \frac{Q_{adv}}{Q_+}, \tag{7}$$

$Q_+$, and $Q_{adv}$ being, respectively, the flux of heat generated by viscous processes, and the advective flux.

For the advective cooling, we shall use an approximate local expression (Abramowicz *et al.*1996[1]),

$$Q_{adv} = Q_+(\frac{H}{r})^2. \tag{8}$$

Therefore,

$$A = 1 - 2Pr. \tag{9}$$

It should be remarked that eq.(7) is equivalent to

$$F_r = Q_+ - Q_{adv}, \tag{10}$$

Where $F_r$ is the radiative cooling. In steady state, the leptons emit exactly the energy they receive from the protons through Coulomb collisional energy exchange, which may be written, approximately, as (Stepney and Guilbert 1981[14])

$$F_r = 4.54 \times 10^{26}\frac{\tau_p^2}{H\theta_e^{1.5}}\left(\theta_p - \theta_e\right)(1+2z)\left(1+\theta_e^{\frac{1}{2}}\right). \tag{11}$$

The energy equation for electrons and pairs comes from the treatment of Comptonization, as given Zdziarski 1985, and it reads

$$\tau = (1+2z)\tau_p = \frac{1+\theta_e^{0.5}}{\theta_e^{0.5}(1+12\theta_e^2)}\ . \tag{12}$$

In writing eq.(12), we have assumed γ=1, which is not very far from observed spectral indexes in a large number of X-ray and γ-ray sources. Substituting eq.(12) into eq.(11), we obtain from the definition of the compactness parameter

$$\ell = \frac{4\pi}{3}\frac{\square_T\,H}{m_e c^3}\ F_r, \tag{13}$$

$$\ell = 0.0504\tau_p\left(\theta_p - \theta_e\right)\frac{(1+\theta_e^{0.5})^2}{\theta_e^2(1+12\theta_e^2)}\ . \tag{14}$$

Now, writing the equation of state for a gas composed of protons, electrons, pairs and radiation, i.e.,

$$P = \frac{m_e}{m_p}(\theta_p + (1+2z)\theta_e + \frac{\ell}{4\pi}(1+\tau_W)\ , \tag{15}$$

Where $m_e$ ,$m_p$ are, respectively, electron and proton masses. $\tau_W$ is the Wien averaged scattering optical depth (Svensson 1984[18]), and it reads

$$\tau_W = \begin{cases} \dfrac{(1+\theta_e^{0.5})}{\theta_e^{0.5}}\dfrac{(1+5\theta_e+0.4\theta_e^2)}{(1+12\theta_e^2)}\ , & \theta_e \le 1 \\ \dfrac{3}{16}\dfrac{(\ln(1.12\theta_e)+0.75)}{\theta_e^{2.5}(1+\theta_e)(1+12\theta_e^2)}, & \theta_e \ge 1. \end{cases} \tag{16}$$

Before we write down the equation for pair production-annihilation equilibrium, we must talk photon balance, from which we can have the radiation field to obtain the rate of pair production by photon-photon interations.

Following Svensson 1984[18], we treat the radiation field as

$$N() = e^{-x}\left(\frac{N_p}{2}\,x^{-2} + \frac{N_W}{2}x^2\right), \tag{17}$$

Where we have specialized for an spectral index of 1. The ratio of intensities is given by Sunyaev and Titarchuk 1980[16], i.e.,

$$\frac{N_W}{N_p} = \frac{\Gamma(1)}{\Gamma(5)}\ P_{sct}, \tag{18}$$

Where Γ is the Gamma function, $P_{sct}$ is the probability of scattering, and $N_p$ is related to the soft input $N_{soft}$ by conservation of the total number of photons.

If $\dot{N}_{soft}$ is the rate of soft photon creation, and we have equilibrium between input of soft photons and output ot photons from the disk,

$$\square_\gamma = \square_{soft}, \tag{19}$$

And

$$N_\gamma = \square_{soft}\frac{H}{c}(1+\tau). \quad (20)$$

Since the photons escape from the disk with an energy~$m_e c^2 \theta_e$ ,

$$N_\gamma = \frac{F_r}{\mathrm{mec^3\theta e}}(1+\tau). \quad (21)$$

However,

$$N_\gamma = \int_{x_0}^{\infty} N(x)\,dx, \quad (22)$$

Where $x_0$ is the soft photon energy in units of $kT_e$ .Therefore,

$$\frac{N_p}{2}\left(\int_{x_0}^{\infty} x^{-2}\, e^{-x} dx + 2\frac{\Gamma(1)}{\Gamma(5)}\tau\right) = \frac{F_r}{\mathrm{mec^3\theta e}}(1+\tau), \quad (23)$$

Where we have used (Zdziarski 1985[20]),

$$P_{sct} = \tau. \quad (24)$$

Now, using eqs.(13), (14), (21), and (24), we may write

$$N_p = \frac{3}{2\pi\sigma_T H}\left(\int_{x_0}^{\infty} x^{-2}\, e^{-x} dx + 0.258\tau\right)^{-1} \times \frac{\ell}{\theta_e}(1+\tau)\,, \quad (25)$$

Which gives for the pair creation-annihilation equilibrium equation (Lightman 1982[8]; Svensson 1982[17])

$$\frac{9}{4\pi^2}(1+\tau)^2\frac{\ell^2}{\theta_e^2} f(x_0,\theta) = B(\theta)\left(\tau^2 - \tau_p^2\right), \quad (26)$$

Where

$$B(\theta) = \frac{\ln{(1.12\theta+1.3)}}{2\theta^2 + \mathrm{n}\,(1.12\theta+1.3)}\,, \quad (27)$$

$$f(x_0,\theta) = 0.693\left(\int_{x_0}^{\infty} x^{-2}\, e^{-x} dx 0.258\tau\right)^{-2} e^{-\frac{2}{\theta}} [\theta^5 + (0.128\tau)^2\,\theta^{-1} + 0.128\tau(\theta+\theta^3)]. \quad (28)$$

## III.THE NUMBER OF CONSISTENT SOLUTIONS TO THE DISK EQUATION

Using eqs.(5), (6), (14) and (15), we finally reduce our system of equations to

$$\frac{\pi}{\sqrt{2}}\frac{m_p}{m_e}A\eta P^{0.5} = \frac{4\square(1+\theta_e^{0.5})^2}{(1+12\theta_e^2)} \frac{(1819.2P\tau_p - \tau_p\theta)(1+12\theta_e^2) - \theta^{0.5}(1+\theta_e^{0.5})}{(1+\tau_W)(1+\theta_e^{0.5})^2 + 249.33\theta^2(1+12\theta_e^2)}, \quad (29)$$

And

$$\frac{9}{8}(\frac{m_p}{m_e})^2(1+\tau)^2\frac{A^2\eta^2 P}{\theta_e^2} f(x_0,\theta) = B(\theta) \times\left(\tau^2 - \frac{1}{8P^2}\left(\frac{\eta}{\alpha}\right)^2\right), \quad (30)$$

Where A, η, τ, f($x_0$,θ), B(θ), $\tau_p$ are given, respectively, by eqs.(4), (7), (12), (15), (26), (27).

For given η, α, $x_0$,r, this is a system of two-coupled equations on the variables P and θ.

We now solve this system of equations specializing for Cygnus X-1, for which we assume canonical values of $M=10M_0$ and $\dot{m}=6.52\times10^{-2}$(Liang and Nolan 1984[7]). We, then, obtain the following solutions for the different sets of parameters.

| $\Theta_e$ | $\Theta_p$ | P | $\tau_p$ |
|---|---|---|---|
| 0.621 | 34.34 | 0.0195 | 0.239 |
| 1.201 | 171.56 | 0.0958 | 0.0485 |
| z | A | H/r | ℓ |
| 0.345 | 0.805 | 0.441 | 0.597 |
| 0.576 | 0.042 | 0.979 | 0.0693 |

Table(1). The solutions to the disk equations for the parameters r=5, $\eta=1.315\times 10^{-3}$,α=0.1, $x_0$=0.1

| $\Theta_e$ | $\Theta_p$ | P | $\tau_p$ |
|---|---|---|---|
| 0.715 | 29.185 | 0.0121 | 0.181 |
| 1.085 | 353.07 | 0.0300 | 0.073 |
| z | A | H/r | ℓ |
| 0.345 | 0.879 | 0.348 | 0.243 |
| 0.385 | 0.700 | 0.547 | 0.304 |

Table(2). The solutions to the disk equations for the parameters r=15, $\eta=6.206\times 10^{-4}$,α=0.1, $x_0$=0.1

| $\Theta_e$ | $\Theta_p$ | P | $\tau_p$ |
|---|---|---|---|
| 0.621 | 34.166 | 0.0194 | 0.240 |
| 1.365 | 84.46 | 0.0987 | 0.0471 |
| z | A | H/r | ℓ |
| 0.342 | 0.806 | 0.440 | 0.597 |
| 0.344 | 0.013 | 0.994 | 0.0231 |

Table(3). The solutions to the disk equations for the parameters r=5, $\eta=1.315\times 10^{-3}$,α=0.1, $x_0$=0.01

| $\Theta_e$ | $\Theta_p$ | P | $\tau_p$ |
|---|---|---|---|
| 0.715 | 29.03 | 0.01207 | 0.182 |
| 1.13 | 420.36 | 0.03107 | 0.0706 |
| z | A | H/r | ℓ |
| 0.341 | 0.879 | 0.347 | 0.242 |
| 0.342 | 0.689 | 0.557 | 0.305 |

Table(4). The solutions to the disk equations for the parameters r=15, $\eta=6.206\times 10^{-4}$,α=0.1, $x_0$=0.01

| $\Theta_e$ | $\Theta_p$ | P | $\tau_p$ |
|---|---|---|---|
| 0.546 | 13.10 | 0.00761 | 0.305 |
| 1.85 | 1518.3 | 0.0956 | 0.0243 |
| z | A | H/r | ℓ |
| 0.341 | 0.924 | 0.276 | 0.428 |
| 0.348 | 0.044 | 0.978 | 0.0719 |

Table(5). The solutions to the disk equations for the parameters r=5, $\eta=1.315\times 10^{-3}$,α=0.2, $x_0$=0.01

| $\Theta_e$ | $\Theta_p$ | P | $\tau_p$ |
|---|---|---|---|
| 0.597 | 7.79 | 0.00425 | 0.258 |
| 1.603 | 2894 | 0.0329 | 0.0334 |
| z | A | H/r | ℓ |
| 0.341 | 0.958 | 0.206 | 0.156 |
| 0.343 | 0.671 | 0.574 | 0.305 |

Table(6). The solutions to the disk equations for the parameters r=15, $\eta=6.206\times 10^{-4}$, $\alpha=0.2$, $x_0=0.01$

The first thing to be remarked is the existence of two solutions. For both them the number of pairs does not outnumber the protons. These solutions are, practically, independent of the energy of the soft photons. Depending on the combination of the parameters the solutions are physically consistent close to the hole. For α=0.1, every solution we found is consistent for r=5. If we go farther out, after a certain radius, the branch which has the greater temperature will become inconsistent, i.e., the proton temperature will be greater than its virial value, or the disk scale height will be greater than the radial distance. We haven't found this radius, but certainly it is inside r=15. For α=0.2 there is only one physically consistent solution: the branch with the greater temperature does not meet consistency. The variables have quite different behaviors in different branches. In the lower branch (the one with the smaller temperature), the electronic temperature decreases and the proton temperature increases as we come close to the hole. In the lower branch, the pressure, the scale height, the Thomson scattering depth, and the compactness increase as we go into the inner region; the radiative efficiency decreases, and the pair number is practically constant. In the upper branch, the pressure, the scale height, the pair number increase as we go closer to the inner radius; the radiative efficiency, the Thomson scattering depth, and the compactness decrease.

## IV.CRITICAL CONDITIONS

We now look for critical conditions for the disk, i.e., parameters values which define the transition from a steady state to a time dependent solution. In order to do so, we start with eq.(29) and treat it as a function of P, assuming all other variables as parameters.

Besides, a cursory analysis reveals that the term $\tau_p\theta_e(1+12\theta_e^2)$ is negligible in most situations. We, then, write eq.(29) as

$$3.22\times 10^2\eta A P^{0.5}=g_0, \quad (31)$$

Where

$$g_0=\frac{643.18\frac{\eta}{\alpha}(1+12\theta_e^2)+\theta_e^{0.5}(1+\theta_e^{0.5})}{(1+\tau_W)(1+\theta_e^{0.5})^2+249.33\theta_e^2(1+12\theta_e^2)}, \quad (32)$$

The left hand side of eq.(31) has a maximum at P=1/6r, which equals 87.8/ √r . Then, in order to have solution , this maximum should be greater or equal to the right hand side of eq.(31), i.e.,

$$\frac{87.8}{\sqrt{r}}\eta\geq g_0. \quad (33)$$

Then, criticality is determined by the value of α: if α is greater than $\alpha_c$ defined by the equation below

$$\alpha_c=\frac{7.35\sqrt{r}(1+\theta_e^{0.5})^2}{(1+\tau_W)(1+\theta_e^{0.5})^2+249.33\theta_e^2(1+12\theta_e^2)}, \quad (34)$$

If α is greater than $\alpha_c$ , the system of equations for the disk will admit two steady solutions, no matter the value of η. However, if $\alpha\leq\alpha_c$ , the system has a critical condition, and it admits two solutions only and if only $\eta_1\leq\eta\leq\eta_2$ , with

$$\eta_1=\frac{\theta_e^{0.5}(1+\theta_e^{0.5})^3\alpha\sqrt{r}}{g_1}, \quad (35)$$

And

$$\eta_2=\frac{2.09\theta B(\theta)^{0.5}\alpha\sqrt{r}}{g_2}, \quad (36)$$

where

$$g_1=7.35(1+\theta_e^{0.5})^2(1+12\theta_e^2)\sqrt{r}-\alpha\,(1+12\theta_e^2)\times[(1+\tau_W)(1+\theta_e^{0.5})^2+249.33\theta_e^2(1+12\theta_e^2)], \quad eq.(37)$$

And

$$g_2=\left[19.739\frac{\theta^2}{\alpha^2}+\frac{1.21\times 10^6}{r}(1+\tau)^2 f(x_0,\theta)\right]^{0.5} \quad (38)$$

To better understand these critical conditions, let us take r=5.

For this value of r, we see that the maximum value $\alpha_c$ can attain is ~2. Therefore, for any α smaller than 2, solutions will be steady outside a region in θ, which borders are determined by the value of α. Let us specialize for α=0.1. The critical region is $3.7954\times10^{-5}\leq\theta\leq0.612$. We now ask if there is a solution for η=0.01 and θ=0.5. For these values of α,θ, and r, we have η outside the range $\eta_1=1.02\times 10^{-4}$ and $\eta_1=5.69\times 10^{-3}$. Using eqs.(4), (35), and (36) we may write

$$\frac{\theta_e^{0.5}(1+\theta_e^{0.5})^3\alpha r^2}{g_1 S(\theta)}\leq \square \leq\frac{2.09\theta B(\theta)^{0.5}\alpha r^2}{g_2 S(\theta)}. \quad (39)$$

## V.CONCLUSIONS

The results we have obtained differ from those of White and Lightman 1989[21], Kusunose and Takahara 1990[10], and Kusunose and Mineshige 1996[11], mainly by the number of solutions to the disk equations and by the conditions under which the system reaches criticality. The critical viscosity parameter $\alpha_c$ , below which the system may develop time behavior, is a function of r and $\theta_e$ . For a given α, we may find a region in $\theta_e$ where $\alpha \leq \alpha_c$. Even under this condition, the system may have two steady state solutions, as long as the accretion rate satisfies $\dot{m}_1 \leq \dot{m} \leq \dot{m}_2$ , where $\dot{m}_1$ and $\dot{m}_2$ are functions of α, r, and $\theta_e$. To make these points more clear, let us take α=0.1 and r=5. We, then, obtain $\alpha_c$=0.1 at $\theta_e$=3.8x10$^{-5}$ and $\theta_e$ =0.612. For 3.8x10$^{-5}$ $\leq \theta_e \leq$0.612, $\alpha_c \geq$0.1. Therefore, for $\theta_e$=0.5, and α=0.1, there will be two steady solutions only and if only 5.06x10$^{-3}$ $\leq \dot{m} \leq$0.282. If we take, at r=5, values of $\theta_e$ , in that range, that maximize $\alpha_c$ ,$\dot{m}_1$, and minimize $\dot{m}_2$ , we obtain $\alpha_c$~2.0, $\dot{m}_1$~0.033, and $\dot{m}_2$~0.2. As compared to White and Lightman 1989[21] results, ours reflect the fact we have included advection in the energy equation and have adopted a different approach to the treatment of Comptonization. It seems that the inclusion of advection in the pair equation, as White and Lightman 1989[21] did, is a minor effect as compared to the inclusion of this term in the energy equation. We have included advection in a rather local manner (Abramowicz *et al.*1996[1]), explicitly dependent on r, and this has introduced an explicit r dependence in our equations, and not only an implicit dependence through η. This somehow helps to explain

Our different result as compared to Kusunose and Mineshige 1996[11] results, besides the fact they have used different radiative cooling, and used another local approach to the advection treatment. Though Kusunose and Takahara 1990[10] have not included advection in their disk with pairs produced by soft external comptonized photons, they obtain a rather similar result to the maximum accretion rate. Their results are, however, very sensitive to the soft photon energy. Our results ,qualitatively ,are in agreement with those of Chen *et al.* 1995[5] ,who have analyzed solutions to the disk equations in terms of $\dot{m}$ and α and have found a critical $\alpha_c$, above which there is always at least one optically thin advection dominated solution for a given accretion rate. For $\alpha \leq \alpha_c$ , there is a maximum accretion rate, above which there are no steady state solutions. We, finally, may summarize the main results obtained in this work, in which we have revisited the inner region of a two temperature soft photon comptonized accretion disk with pairs produced by photon-photon interactions:

- We have found two solutions to the disk equations;
- For both these solutions the number of pairs is always smaller than the number of protons;
- These solutions practically do not depend on the soft photon energy;
- The consistency of the solutions close to the hole depend on the parameters combinations;
- Going farther out, after a certain radius, the upper solution is no longer consistent, i.e., the proton temperature will be greater than its virial value, or the disk scale height will be larger than the radial distance;
- In the lower branch, the one with the smaller temperature, the electronic temperature, the radiative efficiency decrease, the pair number is practically constant, and the proton temperature, the pressure, the scale height, the scattering depth, and the com pactness increase as we come closer to the hole;
- In the upper branch, the electronic temperature, the pressure, the scale height, the pair number increase ; the proton temperature, the radiative efficiency, the Thomson scattering depth, and the compactness decrease as we come closer to the inner radius;
- If α increases, temperature decreases in the lower branch and increases in the upper one;
- If $\dot{m}$ increases, temperature remains practically constant in the lower branch and increases in the upper one
- For α=0.2, only the lower branch solution is consistent;
- In the lower branch, cooling is dominated by radiation and, in the upper branch, cooling is dominated by advection;
- It may happen that at certain r, both solutions are consistent. When we go farther out, just one of these solutions keeps consistency.